\documentstyle[12pt]{article}

\let\useblackboard=\iftrue
\iftrue
\def\nbaselineskip{15pt}
\fi
\useblackboard
\font\blackboard=msbm10 scaled \magstep1
\font\blackboards=msbm7
\font\blackboardss=msbm5
\newfam\black
\textfont\black=\blackboard
\scriptfont\black=\blackboards
\scriptscriptfont\black=\blackboardss
\def\Bbb#1{{\fam\black\relax#1}}
\else
\def\Bbb{\bf}
\fi

\def\sect#1{\subsection{#1}\setcounter{equation}{0}}
\catcode`@=11
\def\@cite#1#2{\if@tempswa [#1]\else$^{\scriptscriptstyle
\mbox{\rm\scriptsize#1}}$\fi}
\newcommand{\eqn}{\begin{eqnarray}}
\newcommand{\enq}{\end{eqnarray}}
\newcommand{\eqa}{\begin{array}}
\newcommand{\ena}{\end{array}}
\newcommand{\eq}{\begin{equation}}
\newcommand{\en}{\end{equation}}

\newcommand{\no}{\nonumber}

\def\comments#1{}
\newcommand{\IZ}{{\Bbb{Z}}}

\def\1N{$1\over N$}

\def\CPN{$\Bbb{C}P^{N-1}$}
\def\Gt{$\Bbb{G}(2,N)$}
\def\G24{$\Bbb{G}(2,4)$}
\def\Go{$\Bbb{G}(1,N)$}
\def\Gkn{$\Bbb{G}(k,N)$}

\def\Gs{$\Bbb{G}(3,6)$}

\def\C{$\Bbb{C}P$}

\def\del{\partial}
\def\half{{1\over 2}}
\def\Tr{{\rm Tr\ }}

\def\ket#1{|#1\rangle}
\def\vev#1{\langle{#1}\rangle}

\begin{document}
\setlength{\unitlength}{0.25cm}
\begin{titlepage}
\hfill{\vbox{\hbox{{\sc OUTP-94}-22P}}}
\vspace{1.5cm}

\begin{center}
{\LARGE
Grassmannian $\sigma$-models and Topological-\\
Antitopological Fusion}\\[0.4in]

{
Mich\`ele Bourdeau
\footnote{{\tt bourdeau@thphys.ox.ac.uk}}}\\
Dept. of Theoretical Physics\\
Oxford University\\
Oxford, OX1 3NP\\[1.5cm]

\end{center}

\vfil
\begin{center}
{\sc Abstract}
\end{center}

\begin{quotation}

We study the topological-antitopological fusion equations for
supersymmetric $\sigma$-models on  Grassmannian manifolds \Gkn.
We find a basis in which the metric becomes diagonal and the $tt^*$
equations become tractable. The solution for the metric of \Gkn\
can then be described in terms of the metric for the \CPN models.
The IR expansion helps clarify the picture of the vacua and gives
the soliton numbers and masses. We also show that the $tt^*$ equation
for \Gkn\ in the large $N$ limit is solvable, for any $k$.

\end{quotation}
\vfill
September 1994 \hfill

\end{titlepage}
\setlength{\baselineskip}{\nbaselineskip}
\newpage

\sect{Introduction}

We investigate two-dimensional supersymmetric
sigma models on Grassmannian target spaces  using techniques
developped in the study of $N=2$ supersymmetric theories in two
dimensions. These techniques follow from the possibility of describing
many models in terms of  Landau-Ginsburg type actions (for a
review, see \cite[x,y]{lvw,warner}). These actions are characterized
by a superpotential  which obeys non-renormalization
 theorems and can be used to study both
conformal and massive theories. The superpotential encodes the
chiral ring of the model under consideration and many of the
properties of the model can be determined through a metric and a new
index computed from the chiral ring.
This metric, defined on the space of Ramond
supersymmetric ground-states (which, by spectral flow, are
in one-to-one correspondence with the chiral superfields),
is the ordinary inner product in the
Hilbert space of states and is determined by the
topological-antitopological differential fusion equations obtained
from the chiral ring.\cite{cv1} The metric can be thought of as a
generalization of the Zamolodchikov metric away from the conformal
point. This metric and the new index derived from it are helpful for
understanding various properties of the model, like the scale and
coupling dependence and the soliton spectrum. The new index can be
obtained from a set of integral
equations  by means of the thermodynamical Bethe ansatz, given
the exact S-matrix. The index being related to the metric, these
integral equations are equivalent to
the $tt^*$ differential equations. However both sets of equations
have proven difficult to solve, and their
equivalence has only be shown numerically for some simple
cases.\cite{cv1,cfiv,cv2}

The differential  equations  simplify when one
considers a model with an infinite number of chiral superfields,
as was done in \cite[x]{bd} for the \CPN model in the large $N$ limit.
 In this case the $tt^*$ equations determining
the metric become an equation first studied in the context of
self-dual gravity which is  related to a symmetry reduction
of Pleba\'nski's ``heavenly'' equation for a self-dual
K\"ahler potential in D=4.
The model is solved using finite temperature results and
methods inspired from self-dual gravity. This example shows
 how  the $tt^*$ formalism contains a lot
of information about a particular quantum field theory
 without having to solve the theory completely.

Here we consider  supersymmetric $\sigma$-models on
Grassmannian target spaces  \Gkn, the \CPN model
being the simplest Grassmannian model \Go.
 These $(1+1)$ dimensional field theories have many analogies with
$(3+1)$ dimensional non-abelian gauge theories: both have instantons,
 are conformally invariant at the classical level, and have
dynamical mass generation and asymptotic freedom at the quantum level.

 We derive the $tt^*$ equations for the ground-state metric
of the Grassmannian $\sigma$-models  \Gkn\  for any $k$ and $N$.
The equations become solvable in terms of  the metric for \CPN
 when written in a basis for which the
metric becomes diagonal. The boundary conditions in the
infra-red and ultra-violet are then easily obtained.
In the IR limit, a clear picture emerges for
the numbers and  masses of solitons interpolating between various vacua
and completes the description in \cite[x]{cv2}.

The $tt^*$ equations for the \Gkn\ models
are then studied  in the large  $N$ limit. We show that the equation
is solvable in this limit for any $k\geq 2$, once the solution for
$k=1$ (the large $N$ \CPN model\cite{bd}) is known.

The Grassmannian models are interesting in other ways, since they
have not been solved completely as quantum
field theories, in particular their exact S-matrices and
spectrum have not been fully determined, as well as their
finite temperature properties. The results reported here should
provide further insight into solving them.

In section 2. we review the Grassmannian $\sigma$-models.
In section 3., we review the fusion equations for the \CPN\ model and
 derive the $tt^*$ equations and their asymptotics for \Gkn. In
section 4. we give some examples, and in section
5. we derive the $tt^*$ equation for large $N$ and fixed $k$, and give
its solution. The concluding remarks are in section 6.

\sect{ Grassmannian $\sigma$-models}

Supersymmetric non-linear $\sigma$-models define maps from spacetime
into a riemannian target manifold $M$. If the target manifold is
K\"ahler, the model will have $N=2$ supersymmetry.\cite{zu}
(For a review, see
refs \cite[x,y,z]{ag,novi,pere}).

\noindent We  study here supersymmetric $\sigma$-models defined on the
complex Grassmannian  manifolds \Gkn. These spaces
have  (complex) dimension $k(N-k)$ and consist of
all $k$ dimensional subspaces
of the complex vector space $V\cong${\bf C}$^N$.

The supersymmetric Ramond ground states of an $N=2$ $\sigma$-model
(and thus, by spectral flow, the chiral superfields) are in
one-to-one correspondence with the complex cohomology classes of the
target space. The chiral ring will be a deformation of the classical
cohomology ring of the manifold, due to instanton
corrections.\cite{wit2} For the \CPN models, the classical cohomology ring is
generated by the K\"ahler form $X$, with relations $X^N=0$. Instanton
corrections modify the ring relations to $X^N=\beta$.

We now describe the classical cohomology of the Grassmannian manifold
\Gkn. Over the Grassmannian \Gkn\ there is a
``tautological'' $k$-plane bundle $E$, whose fiber at each point $x$
in \Gkn\ is the $k$-plane labelled by $x$, and a complementary bundle
$F$ of rank $(N-k)$, producing the exact sequence
\eq\label{es}
0\rightarrow E\rightarrow V\cong {\bf C}^N\rightarrow F\rightarrow 0.
\en
The  classical cohomology of  \Gkn\ is generated from  the Chern classes
$X_i\equiv c_i(E^*)$, with certain relations, where $X_i$ is a
$(i,i)$ form and $E^*$ is the dual of $E$
(see \cite[x,y,w,z]{bott,ken,gep,wit1}).

 Define the following  generating function
\eq\label{gen}
c_t=\sum_{i=0}^kc_i(E^*)t^i=\sum_{i=0}^kX_it^i
\en
In the same way, let $Y_j=c_j(F^*)$, where
 $F^*$ is dual to $F$.

\noindent It follows from (\ref{es}) that
\eq\label{coh}
c_t(E^*)~.~c_t(F^*)=\sum_{i\geq 0}X_it^i~.~\sum_{j\geq 0}Y_jt^j=1
\en
and the classical cohomology is  generated by the $X_i,\,Y_j$ with
conditions  (\ref{coh}).

\noindent In particular they imply
\eq\label{y}
Y_j=0,\qquad\quad{\rm for}\quad N-k+1\leq j\leq N.
\en
The quantum cohomology ring results from a modification to relations
(\ref{coh}) of the form\cite{wit1}
\eq\label{qc}
c_t(E^*)~.~c_t(F^*)= 1+(-1)^{N-k}~ t^N
\en
Then,
\eq\label{ring}
\bigl(\sum_{i=0}^kX_i~t^i\bigr)~.~\bigl(\sum_{j=0}^{N-k}Y_j~t^j\bigr)
=1+(-1)^{N-k}~t^N
\en
and (\ref{y}) gets modified to
\eq
Y_{N+1-i}+(-1)^{N-k}~\delta_{i,1}=0,\quad 1\leq i\leq k
\en
The quantum cohomology ring is generated by
polynomials in the $X_i$'s (if one eliminates for example the $Y_j$'s)
subject to the constraints (\ref{ring}) and its dimension is $N!/k!(N-k)!$.

We are now interested in considering the Grassmannian $\sigma$-models
as quantum field theories and finding the chiral superfields which are in
one-to-one correspondence with the cohomology classes of the target
space, and thus generate the quantum cohomology ring.

A convenient way to find these fields and determine the chiral
ring is through a
Landau-Ginsburg type description for the effective action of the
model under consideration.

Many $N=2$ supersymmetric theories in two dimensions admit a
Landau-Ginsburg description if their superspace Lagrangian
is of the form
\eq\label{lglang}
{\cal L}=\int d^4\theta \sum_i \phi_i\bar{\phi}_i +\int d^2\theta
W(\phi_i)+h.c.
\en
where $\phi_i$, $\bar{\phi_i}$ are the chiral and antichiral
$(a,c)$ superfields and the superpotential $W$  is an
analytic function of the complex superfields which obeys non-renormalization
theorems.
The ground states of the theory are  $dW(\phi)=0$.
The chiral ring is the ring of polynomials generated by the
$\phi_i$
modulo the relations
$dW(\phi)/d\phi_i = D \bar D \phi_i \sim 0$.
(For a review, see \cite[x,y]{lvw,warner}.)

It turns out that the effective action for both the \CPN\ and
\Gkn\ $\sigma$-models have the form of a Landau-Ginsburg
theory:\cite{dadda,cv2,bd}

A manifestly $N=2$ supersymmetric effective Lagrangian
for the \CPN model is
\eq\label{lgform}
   {\cal L} = \int d^4\theta \left [ \sum_{i=1}^N \bar{S}_i
    e^{-V} S_i + {A\over{2\pi}} V\right ]
\en
where $S_i$ are $N$ chiral superfields which become
the homogeneous coordinates on \CPN. Their complex components are the
$N$ complex scalar fields  $n_i$ and fermion
fields $\psi_i$. $V$ is a real vector superfield which contains the
auxiliary fields of the theory. (For a review, see
\cite[x]{bd} and references therein).

As quantum field theories,   the
Grassmannian $\sigma$-models (for $k\geq 2$) can be thought of
as generalizations of the \CPN
models. The Lagrangian has a similar form
\eq\label{lgg}
   {\cal L} = \int d^4\theta \left [ \sum_a \bar{S}_a
    e^{-V} S_a + \alpha \Tr V\right ] .
\en
where now the
chiral fields $S_{ia}$ carry two indices, a `gauge' $U(k)$ index $i$,
and a `flavour' $SU(N)$ index $a$,  since  there are now
$(N\times k)$-matrix scalar fields $n=(n_i^a)$ and
$(N\times k)$-matrix Dirac spinor fields $\psi=(\psi^a_i)$.
$V$ is  a $k\times k$ matrix of superfields with gauge group $U(k)$.
(For a detailed description of Grassmannian $\sigma$-models, see
\cite[x]{abd}.)

 Integrating out the superfields $S_i$ in (\ref{lgform}),
one obtains an effective action for the \CPN models  which has the
form of a Landau-Ginsburg model. By gauge invariance
only the field-strength superfields $X$ and $\bar{X}$ remain since $V$ is
not gauge invariant:
\eqn
S_{\rm eff}={N\over{2\pi}}\!\int\!\!d^2x\left\{
\int \!\! d^2\theta W(X)\!+\!\!
\int\!\! d^2\theta \bar{W}(\bar{X})
\!+\!\!\int\!\! d^4\theta[Z(X,\bar{X},\Delta,\bar{\Delta})]\right\}
\enq
with
\eq
W(X)={1\over{2\pi}} X(\log X^N-N+A(\mu)-i\theta).
\en
$A$ is a renormalized coupling and $\theta$  the
instanton angle,
and
\eq
  X=D_L\bar{D}_R V,\qquad\bar{X}=D_R\bar{D}_L V.
\en
The chiral ring is the powers of $X$ mod $dW=0$, or
\eq
X^N = e^{ -A+i\theta} \equiv \beta.
\en

For  the Grassmannian $\sigma$-models, the
generalization is the following.\cite{cv2} The field-strength
superfields (which we now call $\lambda$, $\bar{\lambda}$)
belong to the adjoint representation of $U(k)$ and
are now gauge covariant.
The gauge-invariant objects of (\ref{lgg})
are now the Ad-invariant polynomials in the
field-strengths $\lambda$, with a ring generated
by the $(a,c)$ superfields
$X_i\quad(i=1,2,\dots,k)$ defined by
\eq
\det [t-\lambda]=t^k+\sum_{j=1}^k(-1)^jt^{k-j}X_j.
\en
However, Cecotti and Vafa observe\cite[x]{cv2} that at the
topological field theory level, the $\lambda$'s and $\bar{\lambda}$'s are
independent of each other and therefore, as matrices,  are
diagonalizable and all their eigenvalues $\lambda_m$ are distinct.
Then, without loss of generality, the
functional determinants in the path integral (with all background
fields  constant and fermions vanishing) are the same as
for the \CPN case, since the full background is now abelian,
 and the superpotential takes the form
\eq\label{lgw}
 W_f(\lambda_1,\lambda_2,\dots,\lambda_k)={1\over {2\pi}}
\sum_{j=1}^k\lambda_j(\log \lambda_j^N-N+A(\mu) -i\theta ),\no\\
\en
which fixes the theory completely.

The gauge-invariant fields are now polynomials in the eigenvalues
$\lambda_m$ of the field-strengths $\lambda$ and are generated by the
elementary symmetric functions
\eq\label{var}
X_i(\lambda )\equiv\sum_{1\leq l_1
< l_2 < \dots l_i\leq k}\lambda_{l_1}\lambda_{l_2}
\dots \lambda_{l_i}\quad (i=1,\dots, k)
\en

The ring relations are
$\lambda_j^N=$ const. and the quantum cohomology  of the
Grassmannian $\sigma$-models will be generated by
the elementary symmetric functions $X_i$'s. As quantum field theories, the
Grassmannian $\sigma$-models \Gkn\ can thus be identified as the
 tensor product of $k$ copies of the \CPN $\sigma$-model
with certain redundant states eliminated.\cite{cv2}

This form of the superpotential is also derived in \cite[x]{wit1}.
There Witten shows, by relating the quantum
cohomology of the Grassmannians to a cubic quantum form in a quantum
gauge theory, that the  Grassmannian $\sigma$-model
reduces at very large distances (where massless particles dominate)
to a topological field theory which is a gauged WZW model.
He shows that it is possible to work in a region of space where the
auxiliary fields and the field-strength superfields which couple to
them in the effective action belong now to a diagonal subgroup $U(1)^k$
of $U(k)$  and  (\ref{lgw}) is then
 the gauged WZW model of $U(1)^k$.

\sect{Topological-antitopological fusion equations}

Here we derive the $tt^*$ equations for the
Grassmannian models \Gkn\ for any $k$ and $N$, starting with \Gt.
We first review the $tt^*$ equations for the \CPN model and their
ground-state metric.\cite{cv3,cv2,bd}

\subsubsection*{\it\underline{tt$^*$ equations for \CPN model}}

The chiral ring for the \CPN model on
a K\"ahler manifold is generated
by a single element $X$, the K\"ahler class, with relation
\eq
X^N=\beta
\en
where $-\ln\beta$ is the action for a holomorphic instanton.\cite{cv3}

As a basis for the chiral ring, we take
\eq\label{chb}
   1,X,X^2,X^3,.....,X^{N-1}.
\en

These fields are in one-to-one correspondence with the Ramond
ground-states $\ket{i}$ of the supersymmetric theory. The inner
product of these ground-states $g_{\bar{i}j}\equiv\vev{\bar{i}|j}$
is the metric we are interested in.

The $tt^*$ equations describe the way in which the metric changes
along the renormalization group flow and are determined by the
following differential equations\cite{cv1}
\eq\label{dif}
  \bar{\partial}_{\bar j}(g\partial_ig^{-1})
  =[C_i,gC^{\dagger}_{\bar j}g^{-1}].
\en
The $C_i$ represent the action on the chiral ring of
the operators corresponding to a perturbation by the couplings.
In the present case, we have one coupling $\beta$, and
the equations are characterized
by a single matrix $C_{\beta}$

\[ C_\beta=-{1\over\beta}\left(\eqa{cccccc}
0 & 1 & 0 & \dots & 0 & 0 \\
0 & 0 & 1 & \dots & 0 & 0 \\
\dots & \dots & \dots & \dots & \dots & \dots \\
\dots & \dots & \dots & \dots & \dots & \dots \\
0 & 0 & 0 & \dots & 0 & 1 \\
\beta & 0 & 0 & \dots & 0 & 0
\ena \right) \]

 The model has a $\IZ_N$ symmetry,
which implies that the metric $g_{{\bar j}i}\equiv\vev{\bar{X}^j|X^i}$
is diagonal. The metric
depends only on $|\beta|^2$ due to chiral charge conservation,  and
the $tt^*$ equations  reduce to the affine $\hat{A}_{N-1}$  Toda
equations\cite{cv1}
\eq\label{tt}
     \partial_z\partial_{z^*}q_i +
          e^{(q_{i+1}-q_i)} -
        e^{(q_i-q_{i-1})}=0
\en
where we have
\eqn
q_i&=&\log\langle\bar i\ket{i} -{{2i-N+1}\over{2N}}\log|\beta|^2\no\\
 z &=& N\beta^{1\over N}\no\\
 q_i&\equiv & q_{i+N}
\enq
One can also define the usual  topological metric, which is the
two-point function $\eta_{ij}$ on the space of states and has
non-vanishing elements
\eq
\eta_{i,N-1-i}=1
\en
The `reality' constraints\cite{cv1} result from a relation between
$\eta$ and $g$. They are
\eq\label{real}
\eta^{-1}g(\eta^{-1}g)^*=1
\en
and imply
\eq
\vev{\bar{i}|i}\vev{\overline{N-1-i}|N-1-i}=1
\en
or
\eq
0=q_i + q_{N-i-1}.
\en

A solution to these equations should be determined by the boundary
conditions near $\beta\sim 0$ and $\beta\sim \infty$.

A semiclassical calculation of the metric in the  UV (small $\beta$)
limit shows that\cite{cv3}
\eq\label{uv}
\vev{\bar{i}|i}={{i!}\over{(N-1-i)!}}[2(-\ln|\beta|-N\gamma)]^{N-1-2i}
\en

In the IR (large $\beta$), the $tt^*$ equations give information about
the solitons of the theory.
In \cite[x]{cv1}, semiclassical considerations are used to
derive the leading order IR expansion for the metric
\eq\label{ir}
q_i\sim {{-2\sin[{{2\pi}\over N}(i+\half)]\exp(-4|z|\sin{\pi\over
N})}\over{\sqrt{8\pi|z|\sin{\pi\over N}}}}
\en

For the \C$^1$ and \C$^2$ cases, the $tt^*$ equations become special
cases of the Painlev\'e  ${\rm I\!I\!I}$ equation, for which the
connection formula between small- and large-$\beta$ asymptotics is
known\cite{cv3}, altough the exact solution is not known. For $N>4$,
the equations have not been studied explicitly.

There exists another basis useful in describing the $tt^*$ asymptotics.
This is the canonical basis.\cite{cv1,cv4,cv2}

This basis diagonalizes the ring $\cal R$
\eq
(C_k)^j_i=\delta_{ki}\delta^j_i
\en
in the sense that one can choose representatives of the chiral ring
$\phi_j$, such that
\eq
\phi_j\ket{i}=\delta_j^i\ket{i}\sqrt{W''(X_i)}.
\en
The canonical vacua are thus eigenstates under multiplication by the
chiral ring.

The large-$\beta$ behavior
for the \CPN metric in the canonical basis can be derived
semiclassically.\cite{cv1,cv2}

Call $\ket{{l}_r}$ the canonical vacua
 at the ${{l_r}}$ critical point,
\eq
\quad X({l}_r)=t^{1\over{N}}\exp [2\pi
i{l}_r/N]\quad ({l}_r=0,1,\dots N).
\en
(where $X$ is the chiral primary  of \CPN ).

Then
\eq\label{can}
\vev{\overline{{l}_s}|{l}_r}\cong
\delta_{sr}-i\, {\rm sign}(r-s)\, {N \choose |r-s|}{1\over \pi}
K_0\left(m_{rs}\beta\right)
+\dots,
\en
 where
\eq
\ m_{rs}=4N|\beta|^{1/N}
\sin\Big({\pi |r-s|\over N}\Big)
\en
is the mass of the soliton connecting the two vacua.\cite{cv3,fi1}

For the \CPN model, it is easy to show that the relation between the
 chiral basis and the canonical basis is just a Fourier
transform
\eq\label{four}
\ket {X^s}={1\over{\sqrt{N}}}\beta^{{2s+1-N}\over{2N}}
\sum_{r=0}^{N-1}e^{{{2\pi ir}\over N}(s+\half)}\ket{{l}_r}
\en

\subsubsection*{\it\underline{The ground-state  metric
for the Grassmannian $\sigma$-model}}

We are interested in finding a suitable basis for the metric of
the Grassmannian $\sigma$-model.
In view of (\ref{lgw},\ref{var}), a basis for the chiral ring will consist of
polynomials $P_r(X_i)$ (for $ r=1,\dots , N!/k!(N-k)!$) in the $X_i$'s.
Then one can determine $C_\beta$ from the ring relations and derive the
$tt^*$ equations from (\ref{dif}). However the equations are difficult
to handle in this form. A more enlightening way to obtain and study
the equations is to use the map (\ref{var}).
The metric can then be defined in terms of the variables $\lambda_m$,
in the following way
\eq\label{ch}
\vev{P_r(X_i)|P_s(X_j)}={1\over{k!}}\vev{\Delta(\lambda)
P_r(X_i(\lambda))|\Delta(\lambda)P_s(X_j(\lambda))}_f
\en
where $\Delta(\lambda)$ is the Vandermonde determinant
and is the Jacobian $J=\det (\del X_i/\del\lambda_j)$ of the transformation
in (\ref{var}).

Each  basis element $P_s(X_j)$ can then be written as a polynomial
$\Delta(\lambda)P_s(X_j(\lambda))$ in
the different $\lambda_m$'s. The metric on the left side of (\ref{ch})
is then  a sum over the products of the metrics
for the \CPN  models, which are
diagonal $\vev{\overline{\lambda_m^k}|\lambda_m^s}=\delta_{ks}
\vev{\overline{\lambda_m^k}|\lambda_m^s}$.

 For example, take the Grassmannian model \Gt. The Chern classes
are $X_1$ and $X_2$, where $X_1$ has (complex) dimension 1
and $X_2$ has (complex) dimension 2.
The dimension of the ring is $N(N-1)/2$ and the ring is generated
by  the $\{X_1,X_2\}$. The ring relations will be determined by
constraints on the $\{X_1,X_2\}$ from (\ref{lgw},\ref{ring}).

\noindent A possible choice of basis is
\eq
{\cal R}=\{1,X_1,X_2,X_1^2,X_1X_2,X_1^3,\dots,X_1^iX_2^j\}
\en
where $(i+2j)\leq {\rm dim}_{\Bbb{C}}$\Gt $\,=2(N-2)$.
\noindent The number of elements
with ${\rm dim}_{\Bbb{C}} |m|$ = number of elements with
${\rm dim}_{\Bbb{C}} |2(N-2) -m|$.

\noindent Then  using (\ref{ch}), with
\eqn
X_1&=&\lambda_1 + \lambda_2\no\\
X_2&=&\lambda_1\lambda_2
\enq
and where the Jacobian of the transformation
 is $\Delta(\lambda)=\lambda_1-\lambda_2$, we have
\eqn
&&\vev{X_1^2|X_2}=\half[\vev{\lambda_1|\lambda_1}
\vev{\lambda_2^2|\lambda_2^2}+\vev{\lambda_2|\lambda_2}
\vev{\lambda_1^2|\lambda_1^2}]\equiv\vev{\lambda |\lambda}
\vev{\lambda^2|\lambda^2}\no\\
&&\vev{X_1^2|X_1^2}=\vev{1|1}\vev{\lambda^3|\lambda^3}
+\vev{\lambda|\lambda}\vev{\lambda^2|\lambda^2}\no\\
&&\vev{X_1X_2|X_1X_2}=\vev{\lambda|\lambda}\vev{\lambda^3|\lambda^3}
\enq
and so forth, where $\vev{\lambda^i|\lambda^i}$ are solutions to the affine
$\hat{A}_{N-1}$ Toda equations.

The ground-state metric is still non-diagonal and complicated. However,
the form of the metric  suggests that one might try to
eliminate  off-diagonal terms  by taking linear combinations of the above
basis elements and finding an orthogonal basis
in which the metric becomes diagonal and simple (i.e. with each
component consisting of a single term). We now show that this is possible.

In \cite[x]{cv1}, the authors mention that the
$tt^*$ equations will have the simplest form in a particular
basis, the `flat coordinates' basis,
characterized by a  two-point function metric $\eta$ which
is independent of the perturbing
parameters of the model and  squares to 1  ($\eta^*=\eta^{-1}=\eta$).
(These coordinates are the ones supplied by conformal
perturbation theory and their chiral algebra,
defined through an effective LG potential,  give
the structure constants $C_{ij}^k$ obtained by conformal perturbation
theory.\cite{dvv,blok,lsw,ken,warner})
We now show that we can find a basis with such an $\eta$, and this choice
makes the $tt^*$ equations tractable.

 Let's first consider  \Gt.
We need  a basis with $N(N-1)/2$ elements.

\noindent Consider  basis elements of the form
$\ket{mn}\equiv\ket{\lambda^m\lambda^n}$ with $n>m$ such that
\eqn
\ket{mn}&=&-\ket{nm}\no\\
        &=&\quad 0 \quad\quad {\rm if}~~ m=n
\enq
We can write
\eqn
\vev{m'n'|mn}&=&\vev{m'|m}\vev{n'|n}
-\vev{m'|n}\vev{n'|m}\no\\
&=&\delta_{mm'}\delta_{nn'}\vev{m'|m}\vev{n'|n}
-\delta_{mn'}\delta_{m'n}\vev{m'|n}\vev{n'|m}
\enq
since we know that the \CPN metric is diagonal.

Then the metric is defined as
\eqn
\vev{mn|mn}&=&\vev{m|m}\vev{n|n}\no\\
           &=&-\vev{mn|nm}\no\\
           &=&\quad\vev{nm|nm}\no\\
           &=&\quad 0 \quad\quad {\rm if}~~ m=n
\enq

Our basis elements for the chiral ring are now the $N(N-1)/2$
elements $\ket{mn}$ with $n>m$.

The reality constraints for \CPN
\eq
\vev{m|m}\vev{N-1-m|N-1-m}=1
\en
become for \Gt\
\eq
\vev{mn|mn}\vev{N-1-m,N-1-n|N-1-m,N-1-n}=1.
\en
The topological metric for the \CPN model is
\eq
\eta_{i,N-1-i}=\vev{X^iX^{N-1-i}}=1.
\en
For the \Gt\ models,  the non-zero elements of
$\eta$ are now
\eq
\eta_{mn, N-1-m\, N-1-n}=\vev{mn, N-1-m\, N-1-n}=1
\en
or, since our basis is defined with $m<n$,
\eq
\vev{mn, N-1-n\, N-1-m}=-1.
\en
This shows that $\eta$ has the form

\[ \eta=\left(\eqa{ccccccc}
0 & 0 & \dots & \dots & 0 & 0 & -1 \\
0 & 0 & \dots & \dots & 0 & -1 & 0 \\
0 & 0 & -1 & \dots & 0 & 0 & 0 \\
0 & 0 & 0 & \dots & 0 & 0 & 0 \\
0 & 0 & 0 & \dots & -1 & 0 & 0 \\
0 & -1 & \dots & \dots & 0 & 0 & 0 \\
-1 & 0 & \dots & \dots & 0 & 0 & 0
\ena \right) \]
where the block in the middle consists of a
diagonal  matrix with elements $-1$,
the size of the block being determined by the
condition that  $m+n=N-1$.

We have $\eta^*=\eta^{-1}=\eta$ and
the reality constraints (\ref{real}) are now
\eq
g\eta g^T=\eta.
\en
The metric $g$ is orthogonal with respect to $\eta $.

The matrix $C_\beta$ characterizing the relations in the chiral ring is
easily determined in this basis. The perturbing operator corresponds to
the chiral superfield
$X_1\equiv \lambda_1 +\lambda_2$
(with a factor of $1\over\beta $ coming from the
action of \CPN ).

\noindent The  algebra of the perturbing operator is then represented by
\eq\label{alg}
X_1\,\,\ket{mn}=\ket{m+1,n}+\ket{m,n+1}
\en
with
\eq
\ket{r,N}=\beta\ket{r,0}=-\beta\ket{0,r}=-\ket{N,r}
\en
since one can think of it as being
\eq
(\ket{\lambda_11}+\ket{1\lambda_2})\otimes\ket{\lambda_1^m\lambda_2^n}=
\ket{\lambda_1^{m+1}\lambda_2^n}+\ket{\lambda_1^m\lambda_2^{n+1}}
\en
The chiral algebra for the other operators can be determined in the
same way, take for example $X_2=\lambda_1\lambda_2$, then
\eq
\ket{\lambda_1\lambda_2}\otimes\ket{\lambda_1^m\lambda_2^n}=
\ket{\lambda_1^{m+1}\lambda_2^{n+1}}
\en
or
\eq
X_2\ket{mn}=\ket{m+1,n+1}
\en
and in general, we have combinations of the following possibilities
\eqn
\ket{\lambda_1^i+\lambda_2^i}\otimes\ket{\lambda_1^m\lambda_2^n}
&=&\ket{\lambda_1^{m+i}\lambda_2^n}+\ket{\lambda_1^m\lambda_2^{n+i}}\no\\
\ket{\lambda_1^k\lambda_2^l}\otimes
\ket{\lambda_1^m\lambda_2^n}&=&\ket{\lambda_1^{m+k}\lambda_2^{n+l}}
\enq
with
\eq
\ket{r,N+s}=\beta\ket{r,s}=-\beta\ket{s,r}=-\ket{N+s,r}
\en

\subsubsection*{\it\underline{The tt$^*$ equations for \Gt }}

 From the form of the metric, we can either use  (\ref{dif})
with $C_\beta$ as defined in  (\ref{alg}), or (\ref{tt}) to find the
$tt^*$ equations. We have
\eqn
g_{\overline{ij},ij}=\vev{\overline{ij}|ij}=\vev{\bar{i}|i}\vev{\bar{j}|j}
\qquad \qquad j>i,\qquad i&=&0,\dots, N-2\no\\
 j&=&1,\dots, N-1
\enq
Defining
\eqn\label{def}
q_{ij}&=&\ln g_{\overline{ij},ij}
-{{(i+j)-N+1}\over{N}}\log|\beta|^2\no\\
 z &=& N\beta^{1\over N}\no\\
 q_{ij}&\equiv & q_{i+Nj}\quad {\rm and}\quad q_{ij}\equiv  q_{ij+N}
\enq
the $tt^*$ equations become for \Gt\
\eq
\del_{\bar{z}}\del_zq_{ij}+e^{q_{ij+1}-q_{ij}}
-e^{q_{ij}-q_{ij-1}}+e^{q_{i+1j}-q_{ij}}-e^{q_{ij}-q_{i-1j}}=0
\en
where an exponential term is ignored if it contains
any $q_{ij}$ with $i=j$ (this will happen if $j=i+1$, then it is easy
to see using (\ref{def}) that the 3rd and 4th  terms on the LHS
cancel each other out).

 The reality constraints can be expressed as
\eq
q_{ij}+q_{N-i-1N-j-1}=0,
\en
which imply
\eq
q_{i,N-i-1}=0,\qquad\quad i=0,\dots,N-2.
\en

We still need  the boundary conditions. For the small-$\beta$ limit,
we use (\ref{uv}) applied to each factor in the metric. For the
large-$\beta$ limit, we first
determine  all one-soliton contributions to $q_i$ for the \CPN models
using (\ref{can}) and (\ref{four}). This will give us the numerical
coefficient in front of the RHS of (\ref{ir}) (which is related to the soliton
numbers) and all the next order
contributions to $q_i$ coming from fundamental solitons.
Then we can get the one-soliton contributions to the asymptotic behavior of
$q_{ij}$.

 In \cite[x]{cv1}, the leading order one-soliton contribution in the
chiral basis (\ref{ir}) is derived from taking the Fourier transform
of the leading
correction (calculated semiclassically) to
$\vev{{\overline{{l}}}_r|{l}_{r+1}}$, the
one-soliton sector of smallest mass. Extending the calculation to include
all one-soliton sectors, we find all fundamental soliton contributions
(coming from the canonical basis) to the IR asymptotics of the metric
for  \CPN
\eq
q_i=\sum_{r=1}^{N-1} {N \choose r}\sin\Big[{{2\pi r}\over
N}(i+\half)\Big]{1\over \pi}
K_0(m_r\beta) +\dots
\en
for $i=0,\dots ,N-1$ and the masses are
\eq
m_r=4N|\beta|^{1\over N}\sin \Big({{\pi r}\over N}\Big).
\en
\noindent The IR boundary conditions for the \Gt\ models then become
\eqn
q_{ij}=q_i+q_j &=&\sum_{r=1}^{N-1} {N \choose r}\Big\{\sin\Big[{{2\pi r}\over
N}(i+\half)\Big]+\sin\Big[{{2\pi r}\over N}(j+\half)\Big]\Big\}{1\over \pi}
K_0(m_{r}\beta) +\dots\no\\
&=&\sum_{r=1}^{N-1} {N \choose r}2\sin\Big[{{\pi r}\over
N}(i+j+1)\Big]\cos\Big[{{\pi r}\over N}(i-j)\Big]{1\over \pi}
K_0(m_{r}\beta) +\dots
\enq
and the leading contribution  is given by the sum of the $r=1$ and
$r=N-1$ terms since the corresponding masses are smallest.

\subsubsection*{\it\underline{ The $tt^*$ equations for \Gkn }}

The $tt^*$ equations easily generalize to \Gkn.
The symmetries of the metric are such that
\eq
\vev{\overline{m_1m_2\dots m_{k}}|m_{i_1}m_{i_2}\dots m_{i_{k}}
}= \epsilon_{m_{i_1}m_{i_2}\dots m_{i_{k}}}
\vev{\overline{m_1m_2\dots m_{k}}|m_1m_2\dots m_{k}}
\en
where $\epsilon_{m_{i_1}m_{i_2}\dots m_{i_{k}}}$ is the totally
antisymmetric tensor. Defining
\eqn
q_{l_1l_2\dots l_k}&=&\ln g_{\overline{l_1l_2\dots l_k},l_1l_2\dots
l_k} - {{2\sum_i l_i -k(N-1)}\over{2N}}\log|\beta|^2\no\\
&=& q_{l_1}+q_{l_2} +\dots + q_{l_k},
\enq
the $tt^*$ equations become
\eq\label{kn}
\del_{\bar{z}}\del_zq_{l_1l_2\dots l_k}+
\sum_{i=1}^k\Big\{\exp[q_{l_1l_2\dots l_{i+1}\dots l_k}-q_{l_1l_2\dots l_k}]-
\exp[q_{l_1l_2\dots l_k}-q_{l_1l_2\dots l_{i-1}\dots l_k}]\Big\}
=0
\en
with appropriate boundary conditions and
where
\eq
0\leq l_1< l_2<\dots<l_i<\dots <l_k\leq N-1,
\en
and  again any exponential containing any $q$ with 2
indices the same is ignored.

\subsubsection*{\it{\underline{The canonical basis for \Gkn }}}

We now show that our basis, when Fourier transformed to the canonical
basis, agrees with the expression derived in
\cite[x]{cv2}. There,  the authors derive a formal expression for the
metric of the \Gkn\ $\sigma$-models in the canonical basis in terms
of the canonical metric for \CPN.
In view of (\ref{lgw}), the Grassmannian $\sigma$-models
\Gkn\ can be identified as the
quotient of a tensor product of $k$ copies of the \CPN\ $\sigma$-
model reduced by the action of the replica symmetry $S_N$, which
eliminates certain states, as explained in \cite[x]{cv2}.

They find then
\eqn\label{metric}
\vev{\overline{\{h_1,h_2,\dots,h_k\}}|\{l_1,l_2,\dots,l_k\}}
&=&{1\over{k!}}\sum_{s,t\in S_k}\sigma(s)\sigma(t)
\prod_{\alpha =1}^{k}\vev{\overline{h_t(\alpha)}|l_s(\alpha)}_\alpha\no\\
&=&\det\nolimits_{{\{\overline{h_\alpha}\},\{l_{\beta}\}}}
\Big[\vev{\overline{h_\alpha}|l_\beta}\Big]
\enq
where $\sigma(s)$ is the signature of the permutation $s$,
$\vev{\bar{h}|l}$ is the $N\times N$ matrix giving the ground
state metric for the \CPN\  $\sigma$-model in a canonical basis, and
$\det_{\{\overline{h_\alpha}\},\{l_{\beta}\}}$ is the determinant of the
$k\times k$ minor obtained by selecting the rows $(h_1,h_2,\dots,h_k)$
and the columns $(l_1,l_2,\dots,l_k)$.

Consider  \Gt, we then see that
\eq\label{basis}
\vev{\overline{m'n'}|{mn}}\!=\!|\beta|^{{2\over N}(m+n-N+1)}\!
\!\!\!\sum_{l_1,l_2,h_1,h_2=0}^{N-1}\!\!{1\over{N^2}}
e^{{{2\pi i}\over N}[(m+\half)l_1+(n+\half)l_2]}
e^{-{{2\pi i}\over N}[(m'+\half)h_1+(n'+\half)h_2]}
\vev{\overline{h_1,h_2}|l_1,l_2}
\en
when we Fourier transform  each component of
the metric separately
\eq
\vev{\overline{h_1,h_2}|l_1,l_2}=\vev{\overline{h_1}|l_1}
\vev{\overline{h_2}|l_2}
-\vev{\overline{h_2}|l_1}\vev{\overline{h_1}|l_2}.
\en
This generalizes to \Gkn.

\sect{Examples}

\subsubsection*{\it\underline {\G24\ } }
We  look first at the simplest non-trivial Grassmannian $\sigma$-model
\G24.

\noindent By using (\ref{ring}), the relations that determine the chiral ring
are the following
\eqn
X_1^3&=&2X_1X_2\no\\
\beta&=&X_2^2-X_1^2X_2
\enq
There are 6 elements in the ring and since cohomology elements are of
even degree, with highest degree  the (real) dimension of \G24,
or 8,  the ring is
\eqn
{\cal R}&=&\{1,X_1,X_2,X_1^2-X_2,X_1X_2,X_2^2\}\no\\
        &\equiv &\{\ket{01},\ket{02},\ket{12},\ket{03},\ket{13},\ket{23}\}
\enq

The matrix $C_\beta$ is defined by  the relations in (\ref{alg})
and takes the form

\[ C_\beta={1\over\beta}\left(\eqa{cccccc}
0 & 1 & 0 & 0 & 0 & 0 \\
0 & 0 & 1 & 1 & 0 & 0 \\
0 & 0 & 0 & 0 & 1 & 0 \\
0 & 0 & 0 & 0 & 1 & 0 \\
-\beta & 0 & 0 & 0 & 0 & 1\\
0 & -\beta & 0 & 0 & 0 & 0
\ena \right) \]

The reality constraints are
\eqn
g_{\overline{01},01}~.~g_{\overline{23},23}&=&1\no\\
g_{\overline{02},02}~.~g_{\overline{13},13}&=&1\no\\
g_{\overline{12},12}^2&=&1\no\\
g_{\overline{03},03}^2&=&1
\enq
where
\eq
g_{\overline{ij},ij}=\vev{\lambda^i|\lambda^i}\vev{\lambda^j|\lambda^j}
\en
The $tt^*$ equations are, with
\eqn
q_{01}&=&\log  g_{\overline{01},01}|\beta|=-q_{32}\no\\
q_{02}&=&\log  g_{\overline{02},02}|\beta|^{\half}=-q_{31}\no\\
q_{03}&=&q_{12}=0
\enq
\eqn
&&\del_{\bar{z}}\del_zq_{01}+e^{q_{02}-q_{01}}
-e^{q_{01}+q_{02}}=0\no\\
&&\del_{\bar{z}}\del_zq_{02}+2e^{-q_{02}}
-e^{q_{02}-q_{01}}-e^{q_{02}+q_{01}}=0
\enq
There are only two unknown functions, since this system has the same
number of unknown functions as \C$^3$.

Let's know consider the behavior of the metric in the IR. From the metric
of the \C$^3$  model, the leading term in the one-soliton sector of
lowest mass is of the form
\eqn
e^{q_{01}}=\vev{01|01}&\cong& |\beta|^{-1}\{1+{4 \choose 1}
{{2\sqrt{2}}\over\pi}K_0(m_1\beta)\}\no\\
e^{q_{02}}=\vev{02|02}&\cong& |\beta|^{-{1\over 2}}\{1+
{4 \choose 2}{2\over\pi}K_0(m_2\beta)\}
\enq
and $\vev{03|03}=\vev{12|12}=1,\qquad q_{23}=-q_{01},\qquad q_{13}=-q_{02}.$

By then going to the canonical basis, one can determine the soliton
numbers between various vacua.

For example, calling $\vev{\overline{h_1,h_2}|k_1,k_2}
=\vev{\overline{l_{h_1h_2}}|l_{k_1k_2}}$,
\eqn
\vev{\overline{l_{02}}|l_{01}}&=&i{4\over \pi}K_0(m_1\beta)
-{24\over{\pi^2}}K_0(m_1\beta)K_0(m_2\beta)+\dots\no\\
\vev{\overline{l_{03}}|l_{01}}&=&i{6\over \pi}K_0(m_2\beta)-{16\over{\pi^2}}
K_0(m_1\beta)K_0(m_1\beta)+\dots
\enq
and so forth.

The soliton structure that emerges is the following.
The Grassmannian $\sigma$-models having the same ring structures as the
Kazama-Suzuki LG  models\cite{kasu} $SU(N)/SU(N-k)\times SU(k)\times
U(1)$ at level 1, perturbed by the most relevant operator, which have been
discussed in \cite[x,y,z]{lvw,lnw,flmw,lw1,lw2},
the geometry of their vacuum images in the $W$-plane
(the complex plane with the values of $W(l_{ij})$ plotted) is similar.

For \G24, the representation in the $W$-plane is the same as for the perturbed
$SU(4)/SU(2)\times SU(2)\times U(1)$ model discussed in
\cite[x]{lnw}. The states
$\ket{02}$ and $\ket{13}$ sit at the top of 2 tetrahedra joined
together by the 4 other vacua. Between these 2 states and each of the
4 other states there are solitons interpolating, of mass $m_1$
and multiplicity 4. There are also solitons of mass $m_2$ and multiplicity
6 between each adjacent vacua of the square formed by the 4 states.
There are no
fundamental solitons linking the 2 vertices ($\ket{02},\ket{13}$).

\subsubsection*{\it\underline{\Gs}}

The ring is
\eqn
{\cal{R}}&=&\{1,X_1,X_2,X_1^2-X_2,X_3,X_1X_2-X_3,-X_1^3+2X_1X_2-X_3,
     X_1^2X_2-X_2^2-X_1X_3,\no\\
     &&X_2^2-X_1X_3,X_1X_3,X_1X_2^2-X_2X_3-X_1^2X_3,
     X_3X_1^2-X_2X_3,X_2X_3,X_2^3+X_3^2-2X_1X_2X_3,\no\\
     &&X_1X_2X_3-X_3^2,
     X_3^2,X_1X_3^2,X_3X_2^2-X_1X_3^2,X_2X_3^2,X_3^3\}\}\no\\
     &=&\{\ket{012},\ket{013},\ket{023},\ket{014},\ket{024},
     \ket{015},\ket{123},\ket{025},\ket{034},\ket{124},\ket{035},\no\\
      &&\ket{125},\ket{134}, \ket{045},\ket{135},\ket{234},\ket{235},
      \ket{145},\ket{245},\ket{345}\}
\enq
The $tt^*$ equations are easily derived from (\ref{kn}).

In the IR, we have for example the following expansions
\eqn\label{asymp}
q_{012}=-q_{345}&=&{6 \choose 1}{4\over\pi}K_0(m_1\beta)+{6\choose 3}
{1\over\pi} K_0(m_3\beta)\no\\
q_{013}=-q_{245}&=&{6 \choose 1}{2\over\pi}K_0(m_1\beta)+{6\choose 2}
{\sqrt{3}\over\pi} K_0(m_2\beta)-{6\choose 3}
{1\over\pi} K_0(m_3\beta)\no\\
q_{024}=-q_{135}&=&{6\choose 3}{3\over\pi} K_0(m_3\beta)
\enq
with $m_1=12|\beta|^{1\over 6},\quad m_2=12\sqrt{3}|\beta|^{1\over
6},\quad m_3=24|\beta|^{1\over 6}$.

We can determine which vacua are joined by
fundamental solitons. The metric in the canonical basis
$\vev{\overline{l_{ijk}}|l_{ijk}}$ can  be obtained by inverse Fourier
transform of (\ref{basis}) or by using (\ref{metric}) where the metric
is given by  a three by three determinant.

For example
\eqn
\vev{\overline{l_{012}}|l_{013}}&=&-i{6\over \pi}K_0(m_1\beta)+\dots\no\\
\vev{\overline{l_{013}}|l_{035}}&=&i{15\over \pi}K_0(m_2\beta)+\dots\no\\
\vev{\overline{l_{012}}|l_{015}}&=&-i{20\over \pi}K_0(m_3\beta)+\dots
\enq
The $W$-plane picture is the following
(see also \cite[x]{flmw}). It consists of
two concentric circles with the smaller
one having half the radii of the larger one (in the ratios $m_3:m_1$),
plus two vacua in the middle. Six vacua are at the vertices of a
regular hexagon on the outer circle and consist of the states
$\{\ket{012}\ket{015}\ket{045}\ket{345}\ket{234}\ket{123}\}$ and there
are 12 vacuum images on the inner circle (since images come in pairs and
are mapped to the same point here) with the two sets of states
$\{\ket{013}\ket{014}\ket{145}\ket{245}\ket{235}\ket{023}\}$ and
$\{\ket{125}\ket{025}\ket{035}\ket{034}\ket{134}\ket{124}\}$.
The hexagons have the same orientation. The two degenerate
vacuum images in the middle are $\{\ket{024}\ket{135}\}$.

We now look for the soliton polytope.\cite{lw1,lw2}
{}From our (\ref{asymp}),  we find a fundamental soliton interpolating
between 2 vacua if any two indices characterizing the two vacua are
equal. For example take the vacua $\ket{012}$. It is on the outer
hexagon and will be
connected only to the two adjacent vacua $\ket{015},\ket{123}$
on the outer hexagon (with solitons of mass $m_3$ and multiplicity 20).
It is also connected to the 6 nearest neighbors on the
inner circle, (the vacuum images being degenerate 2 by 2 on the inner circle)
($\ket{014},\ket{025}$) and ($\ket{023},\ket{124}$) with masses $m_2$
and multiplicity 15 and ($\ket{013},\ket{125}$) with masses $m_1$ and
multiplicity 6. Finally $\ket{012}$ is connected with the vacuum in
the middle $\ket{024}$ with mass $m_3$ and multiplicity 20. In the
inner circle, each of the vacua in the first set
$\{\ket{013}\ket{014}\ket{145}\ket{245}\ket{235}\ket{023}\}$ are
connected to all the other vacua in the first set with masses $m_1,m_2,m_3$
and multiplicities $6,15,20$ and the other set is connected among
itself in the same way. Finally each vacua in the first set (the same
is true for the second set) is connected alternatively
to either one or the other vacuum in the middle, depending on whether
there are any two indices the same.

As a summary then, solitons  fall into multiplets of completely
antisymmetric representations (one column Young tableaux) of $SU(6)$,
and the representations and masses
are determined by how far apart the respective vacua are.

\sect{The large $N$ limit}

We are now interested in checking whether the $tt^*$ equations
become solvable in the large $N$ and $k$ limit, in the same way as they were
for the large $N$ \CPN model.\cite{bd}

We consider first the equations for \Gt\ in the large-$N$
limit. There is an immediate generalization
to \Gkn\ for  any $k$.
We  proceed as  in \cite[x]{bd}.

We assume that the metric  becomes a continuous function
of the two variables $s_1\equiv i/N$, and $s_2\equiv j/N$.

\noindent Redefining (see \cite[x]{bd} for notation)
\eq
q_{ij}={1\over N}\ln\vev{\bar{i}|i}\vev{\bar{j}|j}
+2{{(i+j)}\over{N}}\log|\beta|^2,
\en
the $tt^*$ equations become
\eq
{4\over N}\del_{\bar{\beta}}\del_\beta ~q_{ij}+e^{N(q_{ij+1}-q_{ij})}
-e^{N(q_{ij}-q_{ij-1})}+e^{N(q_{i+1j}-q_{ij})}-e^{N(q_{ij}-q_{i-1j})}=0
\en
with $q_{i+N,j}=q_{ij}$ and $q_{i,j+N}=q_{ij}$.

Then, with $q(s_1,s_2)=\ln g(s_1,s_2)$, we have
\eqn
4\del_{\bar{\beta}}\del_\beta \!\!\!\!&&\!\!\!\!q(s_1,s_2)
=\exp{\Big[N\Big\{q(s_1,s_2+{1\over
N})-q(s_1,s_2)\Big\}\Big]}-\exp {\Big[N\Big\{q(s_1,s_2)-q(s_1,s_2-{1\over
N})\Big\}\Big]}\no\\
&&+\exp {\Big[N\Big\{q(s_1+{1\over N},s_2)-q(s_1,s_2)\Big\}\Big]}-\exp
{\Big[N\Big\{q(s_1,s_2)-q(s_1-{1\over N},s_2)\Big\}\Big]}
\enq
and the equation reduces to
\eq\label{eq}
4\del_{\bar{\beta}}\del_\beta\,q(s_1,s_2)
={\del\over{\del s_1}}\exp{\Big[{{\del q}\over{\del s_1}}\Big]}
+{\del\over{\del s_2}}\exp{\Big[{{\del q}\over{\del s_2}}\Big]}
\en
with general solution
\eq
q(s_1,s_2)=\ln g(s_1,s_2)= \ln [g(s_1)g(s_2)]=q(s_1)+q(s_2)
\en
where $s_2>s_1$, and
where $q(s_i)$ is the solution found in \cite[x]{bd} for the \CPN model
in the large $N$ limit.
The solution is again a $k$-th order product of the metric for the
\CPN model with appropriate configurations eliminated.

The generalization for arbitrary (finite or
infinite)  $k$ is, for  \Gkn\
\eq
\del_{\bar{z}}\del_z\,q(s_1,s_2,\dots,s_k)
=\sum_{i=1}^k{\del\over{\del s_i}}\exp{\Big[{{\del q}\over{\del s_i}}\Big]}
\en
with solution
\eq
q(s_1,s_2,\dots,s_k)=\ln g(s_1,s_2,\dots,s_k)=\ln [g(s_1)g(s_2)\dots
g(s_k)]
\en
with
\eq
s_1<s_2<s_3<\dots <s_k.
\en

We note that there is no particular difficulty in going beyond  $k=1$. This is
consistent with an analysis in \cite[x]{wit1} where the author shows
that the Grassmannian $\sigma$-model simplifies in the limit of $k$
fixed, $N\rightarrow \infty$, and therefore everything can be
calculated in an asymptotic expansion in powers of $1\over N$, by
expansion around a certain extremum.

\sect{Concluding Remarks}

We have shown that the $tt^*$ fusion
equations for the ground-state metric of
supersymmetric $\sigma$-models defined on Grassmannian manifolds \Gkn\
can be determined for any $k$ and $N$. The result follows
from writing the $tt^*$ equations  in a particular orthogonal basis
which makes them tractable in terms of the equations for \CPN.
The solution is a $k$-th order product of the ground-state metric for
the   \CPN\ model, where
appropriate anti-symmetrization and normalization procedures are
carried out. We have computed all fundamental soliton contributions to
the $tt^*$ asymptotics in the IR and, by going to the canonical
basis, have shown that our result
is consistent with that proposed by Cecotti and Vafa\cite{cv2} from
path-integral considerations on the full quantum field theory.

As pointed out by the authors in \cite[x]{cv2}, the Grassmannian
models do not fit in the classification scheme of two-dimensional
$N=2$ supersymmetric field theories which determines  the number
of vacua and solitons between them.
These models have various vacua aligned in the $W$ plane,
yielding monodromy matrices $H$ which do not satisfy the
classification equations, and there is
an ambiguity in asking what the soliton numbers are
(see \cite[x,y]{cv2,zas}). By showing how the $tt^*$ equations for the
Grassmannians are related to the ones for the \CPN models, we were able
to study their behavior in the IR  and make the vacua picture clearer.
The multiplicity and masses of  fundamental solitons
 interpolating between any two given vacua
are easily determined.

We are also able to solve the $tt^*$ equation  for \Gkn\ in the large
$N$ limit for any $k$. The solution can be written  in terms of
the metric for \CPN in the large $N$ limit, which was found in
\cite[x]{bd}, using  finite temperature results and large $N$
techniques, and methods inspired from self-dual gravity.
The Grassmannian $\sigma$-models have not been solved completely as
quantum field theories and our results may offer further insights into
them.

\vspace{0.5cm}

I would like to thank M. Douglas for valuable discussions.
This work is supported by a Glasstone Fellowship.

\goodbreak

\end{document}